\documentclass[aps,superscriptaddress,preprint]{revtex4-1}
\usepackage{amsmath,amssymb,amsfonts,amsthm}
\usepackage{graphicx}
\graphicspath{{images/}}
\usepackage{caption,subcaption,graphicx}
\usepackage{anysize,url}
\usepackage{natbib, setspace}
\usepackage{etoolbox}
\bibpunct[, ]{[}{]}{;}{n}{,}{,}
\usepackage{textcomp}
\usepackage{amssymb,amsmath,bm}
\usepackage{mathrsfs,MnSymbol,ifsym,latexsym,bm}
\usepackage{fancyhdr}
\usepackage{hyperref}
\usepackage{appendix}
\usepackage[Euler]{upgreek}
\usepackage{color}
\usepackage{enumitem}
\renewcommand{\vec}[1]{\ensuremath{\mathbf{#1}}} 
\renewcommand{\d}[2]{\frac{d #1}{d #2}} 
\newcommand{\pd}[2]{\frac{\partial #1}{\partial #2}} 
\newcommand{\pdd}[2]{\frac{\partial^2 #1}{\partial #2^2}} 
\newcommand{\gvec}[1]{\ensuremath{\mbox{\boldmath$ #1 $}}} 
\newcommand{\grad}[1]{\gvec{\nabla} #1} 
\renewcommand{\div}[1]{\gvec{\nabla} \cdot #1} 
\newcommand{\mm}{m_\mathrm{max}}

\begin{document}
\title{Linear stability of an active fluid interface}
\date{\today}
\author{Amarender Nagilla}
\affiliation{IITB-Monash Research Academy, Mumbai, 400076, India}
\author{Ranganathan Prabhakar}
\affiliation{Department of Mechanical \& Aerospace Engineering, Monash University, Clayton,VIC,3800,  Australia\\
email: prabhakar.ranganathan@monash.edu}
\author{Sameer Jadhav}
\affiliation{Department of Chemical Engineering, Indian Institute of Technology Bombay, Mumbai, 400076, India}

\begin{abstract}
Motivated by studies suggesting that the patterns exhibited by the collectively expanding fronts of thin cells during the closing of a wound [Mark et al., Biophys. J., 98:361-370, 2010] and the shapes of single cells crawling on surfaces [Callan-Jones et al., Phys. Rev. Lett., 100:258106, 2008] are due to fingering instabilities, we investigate the stability of actively driven interfaces under Hele-Shaw confinement. An initially radial interface  between a pair of viscous fluids is driven by active agents.  Surface tension and bending rigidity resist deformation of the interface.  A point source at the origin and a distributed source are also included to model the effects of injection or suction, and growth or depletion, respectively.  Linear stability analysis reveals that  for any given initial radius of the interface, there are two key dimensionless driving rates that determine interfacial stability. We discuss stability regimes in a state space of these parameters and their implications for biological systems. An interesting finding is that an actively mobile interface is susceptible to fingering instability irrespective of viscosity contrast.
\end{abstract}

\maketitle

\section{Introduction}
``Active" matter consists of large collectives of orientable particles that propel themselves by consuming free energy available in their environment \cite{schweitzer2007brownian}. A rich variety of self-organized behaviour is observed in such collectives, notably in living systems \cite{vicsek2012collective}. Many features of such pattern-forming behaviour are now being understood by taking a continuum or hydrodynamic perspective, and applying the tools of linear and nonlinear stability analysis \cite{marchetti2013hydrodynamics}. Recent work suggests that an interface  between an active and a passive fluid is unstable \cite{callan2008viscous}.

 Saffman and Taylor \cite{saffman1958penetration} analyzed a flat line interface between a pair of passive incompressible and immiscible Newtonian fluids under confinement in two dimensions. Using Darcy's Law to approximate momentum conservation, with a Young-Laplace  boundary condition for the pressure jump across the interface due to surface tension, they showed applying linear stability analysis that a fingering instability arises whenever a fluid is driven into another with a higher viscosity. Paterson later generalized this result to an initially circular interface of arbitrary radius \cite{paterson1981radial}. A vast body of analytical, simulation and experimental work now exists on fingering instabilities in passive Newtonian and non-Newtonian fluids, but the effect of self-propulsive activity at an interface has not been considered in detail.

Callan-Jones et al. \cite{callan2008viscous} showed that spontaneous shape transitions in lamellar cell fragments may be triggered by a fingering instability. In such fragments, actin is continuously polymerized at the edge giving rise to an active driving of the interface. At the same time however, the polymerized actin filaments are broken down in the cell interior such that the overall two-dimensional area of the cell fragment is conserved. Callan-Jones et al. adapted the Saffman-Taylor framework to analyze an initially circular interface with surface tension driven by an active velocity but with a distributed internal sink that ensured area conservation. The bulk fluid within the cell fragment was assumed to be unstructured Newtonian fluid with an effective viscosity. The effect of interfacial bending resistance was ignored. 

Kabaso et al. \cite{kabaso2011theoretical} used an interface-evolution model  to suggest that characteristic cell shapes during cytoskeleton-driven surface migration are the result of interfacial instabilites. A similar model was used by Mark et al. \cite{mark2010physical} to argue that patterns observed at the edge of epithelial tissue during wound healing are the result of a Saffman-Taylor-like instability driven by internal propulsive forces in cells.  In such models, the evolution of the interface at any point is determined by the curvature at that point and its derivatives: the dynamics are entirely local. In contrast, the actual mechanism behind fingering is non-local since the behavior at a point in the interface is in fact coupled to the motion of all other points on the interface through the pressure and velocity fields on either side of the interface. Despite this fundamental difference, however, it has been shown that curvature-evolution models can qualitatively reproduce many of the morphological features of observed in unstable interfaces \cite{brower1983geo}. 

Motivated by these studies, we present here the linear stability analysis of an initially circular, fluid-fluid line interface that is actively driven under Hele-Shaw confinement. We focus here only on the effect of activity at the interface, and neglect any influence of activity in the bulk fluid either inside or outside. We also aim to explore the common features that active interfacial instabilities share with the well known Saffman-Taylor instability. To this end, we also include two additional drivers of interfacial motion, (a) a conventional point source at the origin to model injection of fluid and (b) a distributed source to model biological growth. The following section presents the governing equations and boundary conditions for an interface with surface tension and bending resistance.  Section~\ref{s:LSA } presents the linear stability analysis of such an interface. The conditions for stability and the different kinds of instabilities are discussed in Sec.~\ref{s:results}, along with their implications for non-biological and biological systems. Section~\ref{s:conclusions} summarizes the principal conclusions of this study.

\section{\label{s:model} Model equations}
We consider two immiscible and incompressible fluids in two dimensions under Hele-Shaw confinement. Fluid 1   is surrounded by Fluid 2. The two fluids are  separated initially by a circular interface of radius $R_0$. The inner fluid includes a point  source  of strength $q$  located at the  origin, as well as a uniformly distributed source of strength $\mu$. This distributed source could model exponential growth in  biological systems. A negative value of $\mu$ accounts for depletion in the bulk of the inner domain. For example, in lamellar cell fragments, a distributed depletion  can account for the rate of actin depolymerization within the cell interior \citep{callan2008viscous}.

While the outer fluid is Newtonian, Fluid 1 also contains self-propelled particles. A full continuum description of such a fluid must in principle account for the coupled dynamics of orientational order of the particles. Such continuum descriptions have been reviewed recently by \citet{marchetti2013hydrodynamics}.  However, here we  study the influence of activity at the interface in isolation from any complex behavior in the bulk. Fluid 1 is thus assumed to effectively act as a viscous Newtonian fluid, with the influence of additional active stresses having completely relaxed  at large time scales \citep{callan2008viscous}. This implicitly also assumes  that over the longer time scale of significant interfacial motion, the inner fluid has uniform density in the bulk. Conservation of mass leads to  
\begin{align}\label{eq:continuity}
\div \vec{v}_i \, =\,\begin{cases}
\mu \,  +\,q\,\delta(\vec r)\,, \quad{i = 1}\,;\\
0 \quad{i = 2}\,,
\end{cases}
\end{align}
where $\vec{v}_i$ is the mid-plane velocity field between the confining  Hele-Shaw  surfaces. Inertial effects are further assumed to be negligible, and Darcy's Law is assumed to describe well the momentum of either fluid at the mid-plane \cite{wooding1976multiphase}:
\begin{gather}\label{eq:darcy}
\eta_i \vec{v}_i= -\grad p_i,  
\end{gather}
where $\eta_i$  is the resistance coefficient of the $i$-th fluid. These coefficients are proportional to fluid viscosities \citep{saffman1958penetration}.  

Thus, influence of activity enters our model solely through the boundary conditions. The shape of the interface at any instant of time $t$ is described by vector function $\vec{R}(s,t) $ which gives the position vector at a location $s$ along its contour.  The contour variable $s$  is defined  such that the unit outward normal vector to Fluid 1 is given at any point on the interface by 
\begin{gather}
\vec n = \vec e_z \times \pd{\vec R}{s}\,,
\end{gather}
where $\vec{e}_z$ is the unit normal vector pointing out of the plane of the Hele-Shaw cell. The local curvature vector is
\begin{gather}
\vec H = \pdd{\vec R}{s}
\end{gather}
and the signed curvature $\mathscr H = \vec H \cdot \vec n$ is positive when the interface is concave-outward.

The first effect of activity at the interface is the additional normal velocity imparted to the interface by propulsive forces exerted by active agents. Callan-Jones et al.\citep{callan2008viscous} and Blanch-Mercader et al. \citep{blanch2013spontaneous} modeled propulsion of the cell membrane in lamellar fragments by actin polymerization by an extra active velocity,  $v_a$, above the normal velocity of the inner (Newtonian) fluid at the interface. In general, we can imagine that active agents acting at the interface serve to boost the boundary velocity of Fluid 1.  The interface is itself identified as the inner edge of the outer Fluid 2 so that the interfacial  normal velocity $v_I \,= \,\vec{v}_2 \cdot \vec n$.  Due to the action of the active agents, $v_I = \vec{v}_1 \cdot \vec n + v_a$. Therefore, the kinematic boundary condition for an active interface is a normal velocity jump across the interface:
\begin{gather}\label{eq:b1a}
\left.(\vec{v}_2 - \vec{v}_1) \cdot \vec n  \right|_{\vec{r}\,=\,\vec R} =  v_a \,.
 \end{gather} 
This reduces to the standard kinematic condition for passive interfaces when $v_a = 0$. For an active interface, $v_a$ must be either given as a constant parameter, or must be modeled further. Far away from the interface, $\vec{v}_2 \rightarrow \vec{0}$ as $r \rightarrow \infty$.

We model the interface itself as one with interfacial tension and curvature resistance. The standard pressure-jump boundary condition for such an interface is modified to account for propulsive forces exerted by active agents on the inner side of the interface:  
\begin{gather}\label{eq:b2}
 (p_2 - p_1)_{\vec{r} = \vec{R}} = \gamma \mathscr H -\kappa\left[{\pdd{ \mathscr H}{s} + \frac{1}{2} \mathscr H^3}\right ] + p_a \,,
\end{gather}
where the active pressure $p_a$ represents the propulsive forces exerted on the interface by the inner fluid, and $\gamma$ and $\kappa$ are the interfacial tension and bending rigidity of the interface, respectively.  

The first couple of terms on the right-hand side of the equation above are obtained from the Helfrich free-energy functional  \cite{helfrich1973elastic, zhong1989bending, kabaso2011theoretical, mark2010physical}. As with bulk fluid properties, this represents a highly simplified view of interfaces in active systems, particularly in biology. In classical fluid-fluid interfaces, surface tension is the sole thermodynamic property that characterizes the interface. The behaviour of surfactant-laden interfaces or lipid-bilayer membranes as in cells is more complex. Their interfacial rheology and its influence on instabilities are still under active investigation \cite{sagis2011dynamic}. The tension and bending terms  above obtained through the Helfrich-Canham energy functional represent the simplest model for the elastic (or equilibrium) response of such interfaces \citep{seifert1997configurations}. This simple description further assumes that the interface is chemically homogeneous. In contrast to these molecular interfaces, in systems such as the advancing  of a migrating bacterial colony \citep{gloag2013self}, the interface is a dense layer of micron-sized, rod-shaped particles. The mechanical behaviour of such particle-laden or granular interfaces is not well understood; in their case as well, the model above at best accounts for just the simplest terms arising from  the elastic contribution of the interfacial mechanical response. As such, the analysis presented here represents but a first step in understanding interfacial instability when these complex interfaces are driven actively.

It is of interest to ask whether the driving provided by the active velocity and pressure  at the interface injects net momentum into the fluid system. Marchetti et al.\citep{marchetti2013hydrodynamics} suggested that active matter could in general be classified as either ``dry" or ``wet" systems. Dry systems are those that continuously exchange momentum with a substrate or their environment. In (inertialess) wet systems on the other hand, motion is induced by activity  without any net momentum gain anywhere within the bulk of the active material. In the model here,  momentum contributed by the active velocity $v_a$ is negligible when the Reynolds number is negligibly small. The only injection of momentum due to activity at the interface occurs through the active pressure $p_a$. The net rate of momentum transferred across the interface is therefore $\oint \,p_a \vec{n} \,ds$. If $p_a$ is a constant, this momentum influx is zero irrespective of the shape of the interface. The inner fluid is then a wet system; constant $p_a$ and $v_a$ drive interfacial motion in a force-free manner. 

In their analysis of interfacial instability in lamellar cell fragments, \citet{callan2008viscous} neglected bulk active stresses, and neglected $p_a$.  The interface was driven purely by an active velocity generated by actin polymerization at the interface. No other forces arising from interactions with the substrate were considered. Their model can thus be classified as a wet system.  In contrast, \citet{mark2010physical}  assumed that the interface at the edge of a wound in epithelial cells is forced by an active pressure that depends on the local curvature. When $p_a$ varies across the interface, $\oint \,p_a \vec{n} \,ds$ is in general non-zero, and there can be net injection of momentum into the fluid layer. The physical source of this momentum in systems of motile particles or cells is the interaction of the interface region with the substrate (for instance, forces generated through binding and unbinding of pili in bacteria or focal adhesions in eucaryotic cells). If $p_a$ depends linearly on $\mathscr H$ or  $\partial_s^2 \mathscr H$, those contributions could be absorbed into the corresponding terms in the pressure-jump condition in eqn.~\eqref{eq:b2}. Further, since $p_a$ can be expressed as the derivative of a translation-invariant effective energy functional, $\oint \,p_a \vec{n} \,ds = 0$, and the interface is wet. However, in the model of \citet{mark2010physical}, active forces are exerted only when $\mathscr H < 0$. The interface then is dry, and the active pressure distribution can inject net momentum into the fluid system.

We restrict our attention here to the simplest case where both $v_a$ and $p_a$ are constant. While the analysis is valid for wet interfaces, the qualitative understanding we develop could also provide insight into the behaviour of dry systems with renormalized surface tension and bending resistance coefficients.  Even in more complex systems where $p_a$ and $v_a$ may depend on other variables defined on the interface \citep{kabaso2011theoretical, marchetti2013hydrodynamics}, it is important to first understand the simplest case where these parameters are held constant.

\section{\label{s:LSA } Linear stability analysis}

\begin{figure}[h]\label{Top view}
\includegraphics[width=5cm]{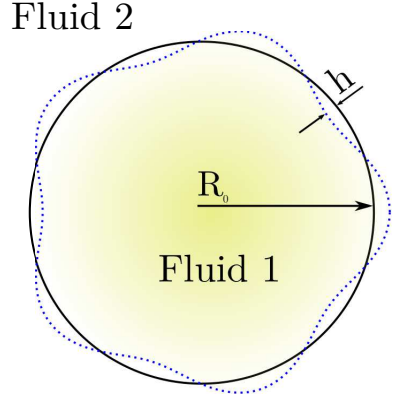}
\centering
\caption{Top view of the inner  and outer fluids in the Hele-Shaw cell. The black  circle represents the initial interface and the blue dotted contour represents the perturbed interface.}
\label{fig:basefig}
\end{figure}

We consider an initially  circular interface of radius $R_0$, as shown in the Fig. \ref{fig:basefig}. Mass conservation along with the normal velocity boundary condition  (eqn. \eqref{eq:b1a}) imply that, for this base-state, 
\begin{gather}\label{eq:vjump}
  v_{I,0}   = \d{R_0}{t}=\frac{\mu R_0}{2}+\frac{q}{2\pi R_0}+v_a= v_{\mu}+v_q+v_a,  
\end{gather} 
where $v_q=q/2\pi R_0$ and $v_{\mu}= \mu R_0/2$ are the interfacial velocity contributions due to the point source $q$ and the distributed source $\mu$, respectively. Solving the governing equations \eqref{eq:continuity} and \eqref{eq:darcy} with the  boundary conditions in eqns. \eqref{eq:b1a} and \eqref{eq:b2} \citep{saffman1958penetration, paterson1981radial}, the pressure distribution in the base-state is obtained as:
\begin{subequations}
\label{eq:P1P2}
\begin{gather}
  p_{1,0} =\eta_1  R_0 \, \left[\frac{v_\mu}{2} \left\{1-{\left(\frac{r}{R_0}\right)}^2\right\} \right]- \,\eta_1  R_0 v_q\,\,\ln{\frac{r}{R_0}},\, \\ 
  p_{2,0}=-\eta_2\, R_0\, \left[{v_\mu}+ v_q +v_a \right]\,\ln {\frac{r}{R_0}} -\frac{\gamma}{R_0}+\frac{\kappa}{2\,R_0^3} + p_{a} \,.
  \end{gather}
\end{subequations}
The corresponding radial velocity fields are:
\begin{subequations}
\label{eq:vr1vr2}
\begin{gather}
v_{r,\,1,\,0} = \frac{\mu \, r }{2} + \frac{q}{2 \pi r} \,, \\
v_{r,\,2,\,0} = \left[ \frac{\mu \, R_0 }{2} + \frac{q}{2 \pi R_0} + v_a \right] \, \frac{R_0}{r} \,.
\end{gather}
\end{subequations}

A small radial perturbation of amplitude $a$,
\begin{gather}
h(\theta, t) \,=\, a \,e^{i m\theta + \omega t} \,,
\end{gather}
is imposed on the initially circular interface. Here, $\theta$ is the polar angle,  $m$ is the azimuthal mode number of the perturbation, and $\omega$ is its growth rate.   The perturbed pressure and velocity fields are  $p_i = p_{i,\,0}  + \hat p_i$, and $\vec{v}_i = \vec{v}_{i,\,0} + \hat{\vec{v}}_i$. From eqns. \eqref{eq:continuity} and \eqref{eq:darcy}, and the solution for the base-state, the perturbations $\hat p_i$ and  $\hat{\vec{v}}_i$  induced by the disturbance at the interface must satisfy,
 \begin{gather}\label{eq:vp}
 \div \hat {\vec v}_i =0 \,, \\ 
 \label{eq:dpp}
 \eta_i  \hat {\vec v}_i = -\grad \hat {p}_i \,.
 \end{gather}
Therefore, $\hat {p}_i$ is harmonic, satisfying
 \begin{gather}\label{eq:phat}
 \grad^2 \hat {p_i}= 0 \,.
 \end{gather} 
 It can be shown that \citep{paterson1981radial}
 \begin{gather}
 \hat{p}_1 \, = \, P_1 r^m h \, \text{ and }  \hat{p}_2 \, = \, P_2  r^{-m} h \,,
 \label{eq:ppertub}
 \end{gather}
satisfy Laplace's equation for the pressure perturbations and the boundary conditions far away from the interface: $\hat{p}_1 \rightarrow 0$ as $r \rightarrow 0$ and $\hat{p}_1 \rightarrow 0$ as $r \rightarrow \infty$. The constants $P_1$ and $P_2$ are determined by applying the boundary conditions at the interface.

The perturbed interface is at
\begin{gather}
\vec R = (R_0 + h)\,\left. \vec{e}_r\right|_{\vec{r} = \vec{R}} \,.
\end{gather}
When the amplitude $a \ll R_0$,  the  unit normal at the interface,
\begin{gather}
\vec n = \left. \vec{e}_r\right|_{\vec{r} = \vec{R}} + \mathscr {O} (a/R_0)\, \left. \vec{e}_\theta\right|_{\vec{r} = \vec{R}}\,.
\end{gather}
Therefore,  correct to first order in $a/R_0$, the normal velocity at the interface is, 
 \begin{gather}\label{eq:vI2}
  v_I =\d {\vec R}{t} \cdot \vec n = \left \{\d{R_0}{t}+\d{h}{t} \right \} =  v_{I,0} \,+\,a\,\omega \,e^{i k\theta + \omega t}\,  = v_{I,0}\,+\,\omega\, h\,.
 \end{gather}
As discussed before, the normal-velocity boundary conditions for an active interface mean that $v_I = \vec{v}_1 \cdot \vec{n} + v_a =  \vec{v}_2 \cdot \vec{n}$. Using Darcy's equation  for $\vec{v}_i$ along with eqn.~\eqref{eq:vjump} for $v_{I,\,0}$ and  the kinematic conditions, we obtain:
\begin{gather}
  v_I  =  v_{\mu}+v_q+v_a + \omega h \, =\, -\frac{1}{\eta_1} \, \left. \left( \pd{p_{1,\,0}}{r} + \pd{\hat {p_1}}{r} \right) \right|_{r=R_0+h} + v_a = -\frac{1}{\eta_2} \, \left. \left( \pd{p_{2,\,0}}{r} + \pd{\hat {p_2}}{r} \right) \right|_{r=R_0+h} \,.
 \end{gather}
Although the base-state pressure and velocity fields $p_{i,\,0}$ and $\vec{v}_{i,\,0}$ have been derived assuming the interface is at $r = R_0$, these functions remain mathematically well-behaved even if $ r > R_0$ for fluid 1 or if $r < R_0$ for fluid 2. Therefore, there is no formal difficulty in evaluating at $r = R_0 + h$ the derivatives of the base-state pressure fields given in eqns.~\eqref{eq:P1P2}. Substituting the general solutions for the pressure perturbations from eqn.~\eqref{eq:ppertub}, and eliminating the constants $P_1$ and $P_2$, we obtain the following fields $\hat{p}_i$ that ensure that the \textit{total} pressure and velocity fields are consistent with the governing equations and the  velocity-jump boundary condition:
\begin{subequations}
\begin{gather}\label{eq:p1hp2h-1}
\hat{p}_1 \,= \,\eta_1\, \left[v_\mu -v_q -\omega R_0\right] \left (\frac{r}{R_0}\right )^{m}\,\left(\frac{h}{m} \right) \,,\\
\hat{p}_2 \,=\, \eta_2\, \left[v_\mu+v_q+v_a+\omega R_0\right] \left (\frac{r}{R_0}\right )^{-m}\,\left(\frac{h}{m} \right) \,.
\end{gather}
\end{subequations}
Additionally, the total pressure fields  should also satisfy the pressure-jump boundary condition  (eqn.~\eqref{eq:b2}) at the interface. Expanding the curvature and its derivatives as a Taylor's series in $h/R_0$, and then applying the pressure-jump boundary condition in eqn.~\eqref{eq:b2} along with the base-state pressures in eqn.~\eqref{eq:P1P2} and the perturbation pressures above, we obtain the following condition -- the dispersion relation --- that the growth rate must satisfy:
\begin{align}
 \begin{split}\label{eq:omegare}
 \omega (m) \, R_0 &= \, \left(\frac{\Delta \eta }{\bar \eta}\,m-1\right)\,v_q\,+\,\frac{ \Delta \eta}{\bar \eta} \, (m-1)\,v_{\mu} \,+\,\left(\frac{\Delta \eta}{\bar \eta}+1\right)\,(m-1) \,\frac{ v_a}{2}  \\
&\qquad  - m\,(m^2-1) \,\frac{  \gamma}{2\,\bar {\eta}\,R_0^2}\,-\,m\,(m^4-\frac{5\,m^2}{2} +\frac{3}{2} )\,\frac{  \kappa}{2\,\bar {\eta}\,R_0^4}\,,
\end{split}
\end{align}
where $\bar \eta =(\eta_1+\eta_2)/2$, $\Delta \eta = (\eta_2 - \eta_1)/2$, and $\Delta \eta/\bar \eta$ is the Atwood number of the mobility. Linear stability of the  $m$-th mode of a perturbation depends on the sign of $\omega$. We discuss next the contributions of different driving parameters on interfacial stability. 
 
\section{\label{s:results} Results and discussion}

\subsection{Dispersion relation and stability states}
For an inelastic ($\gamma$, $\kappa = 0$), passive ($v_a$, $p_a = 0$) interface driven purely by a point source at the origin ( $v_q > 0$; $v_\mu=0$), eqn. \eqref{eq:omegare} yields the well known result  \citep{saffman1958penetration} that 
\begin{gather}
\omega(k) \,=\, \left(\frac{\Delta \eta\, k}{\bar \eta}-\frac{1}{R_0}\right) \, v_q \,,
\end{gather}
 where $k = m/R_0$ is the wavenumber corresponding to the $m$-th mode. In this case, an interface is unstable ($\omega > 0$) as long as $\Delta \eta > 0$. Thus, all wavenumbers above the critical value $k_c = ({\bar \eta}/{\Delta \eta} ) R_0^{-1}$ (or $m_c = {\bar \eta}/{\Delta \eta}$) are unstable.  In the case of a distributed source ( $v_q = 0$; $v_\mu > 0$),
 \begin{gather}
\omega(k) \,=\, \frac{\Delta \eta}{\bar \eta}\, \left(k-\frac{1}{R_0}\right) \, v_{\mu} \,.
\end{gather}
In this case, the critical mode number $m_c = 1$ is  independent of the viscosity. Since  $m \geq 1$ for any physically realizable mode, perturbations of all observable wavenumbers are unstable if the inner fluid growing with a rate $\mu$ is less viscous than the outer fluid. However, in the limit of flat interfaces ($R_0 \rightarrow \infty$), interfacial stability becomes independent of whether the source is a point source or is distributed; it depends only on  $\Delta \eta/ \bar \eta$ and the normal velocity at which the interface is driven. 

In contrast,  for an active interface driven solely by $v_a > 0$, all physically realizable modes are \textit{always} unstable, irrespective of the viscosities of the two fluids, since
 \begin{gather}
\omega(k) \,=\, \left(\frac{\Delta \eta}{\bar \eta} + 1 \right)\, \left(k-\frac{1}{R_0}\right) \, \frac{v_a}{2} \,.
\end{gather}
and $\Delta \eta/ \bar \eta + 1 = \eta_2/\bar \eta$ is always positive.   The terms due to surface tension and bending rigidity in eqn.\eqref{eq:omegare} can be expected to stabilize the interface as their contributions are negative for all $m > 1$. \textit{A constant active pressure $p_a$ further plays no role in the stability of small-amplitude perturbations.} 

We henceforth consider elastic interfaces with non-zero $\gamma$ and $\kappa$. All variables  are rescaled  using the characteristic length scale,  $\ell_c= \sqrt{\kappa/\gamma}$, the time scale $t_c\,=\, (\bar \eta /\gamma) \,  ( \kappa/ \gamma )^{3/2}$, and a characteristic pressure scale, $p_c = \sqrt{{\gamma^3} /{\kappa}}$. We retain the same notation as before for rescaled variables. The rescaled dispersion relation is:
\begin{gather}\label{eq:omega(m)}
\omega(m) = -\,\alpha \,+\,\left( \alpha - \beta + \frac{1}{2\,R_0^3}-\frac{3}{4\,R_0^5}  \right)\,m \, -\,\left(\frac{1}{2\,R_0^3}-\frac{5}{4\,R_0^5}\right) \,m^3\,-\,\left(\frac{1}{2\,R_0^5} \right) \,m^5 \,.
\end{gather} 
The coefficients of the quintic polynomial in $m$ on the right-hand side depend on the dimensionless radius $R_0$ of the base-state, and two dimensionless rates, $\alpha$ and $\beta$, whose definitions in terms of the dimensional model parameters are:
\begin{gather}\label{eq:alpha}
\alpha\,=\, \left(\frac{ \Delta \eta  }{\bar \eta } \right)\,\frac{v_{\mu}}{R_0}\,+\,\frac{v_q}{R_0}\,+\,\left(\frac{\Delta \eta}{\bar \eta}+1\right)\, \frac{ v_a}{2\, R_0  } \,; \qquad
\beta\,=\,\left(1-\frac{\Delta\eta} {\bar {\eta}}\right) \,\frac{v_q}{ R_0}\,.
\end{gather}
With the definitions above, $\beta$ depends only on the injection rate, whereas $\alpha$ depends on all three driving parameters. In addition, since $1 - \Delta \eta / \bar \eta \,= \,\eta_1/ \bar \eta  \, \geq \,0$ always, $\beta$ is positive for a point source and negative for a sink. 

Only integer values of $m$ are physically realizable.  For the purpose of analysis however, we consider $m$ to  be a continuous variable with the region of interest being $ m \geq 1$. We note that,  irrespective of either $\alpha$ or $R_0$,
\begin{gather}
\omega( m = 1) \,=\,-\beta \,.
\label{eq:omega_m_1}
\end{gather}
A small amplitude perturbation with  $m = 1$ corresponds to a small translation of the circular base-state. 
When $\beta = 0$, $\omega ( m = 1) = 0$; that is, the translational mode is marginal in the absence of any injection. This is because of the invariance of the governing equations and the boundary conditions with respect to a translation of the origin when no point source is located at the origin. For systems with a point source (sink) at the origin, the translational mode is linearly stable (unstable) to small perturbations.

As $m \rightarrow \infty$, the growth rate is negative, and decreases asymptotically as $ \omega \sim - m^{5}$, for any $R_0$, $\alpha$ or $\beta$. An extremum in $\omega$ occurs  when
\begin{gather}\label{eq:dmm}
\d{\,\omega}{\,m} \,= \,\alpha -\beta + \frac{1}{2 \,R_0^3}-\frac{3}{4\, R_0^5} \,-\,  \left(\frac{3}{2\,R_0^3}-\frac{15}{4\,R_0^5} \right) \, m^2 \, -\, \frac{5\, m^4}{2\,R_0^5} \,= \,0.
\end{gather}
Conditions for obtaining a real positive root for $m^2$ depend on whether $R_0^2$ is larger or smaller than $5/2$. We are interested in cases where the size of the inner domain is large compared to the length scale arising from the elastic rigidity of the interface. We therefore only present below the analysis for $R_0^2 > 5/2$. Under such conditions, surface tension or bending rigidity stabilize the interface at all modes. Although a constant $p_a$ does not affect the growth of small perturbations, as noted earlier, when $p_a$ depends on local curvature,  surface tension or bending rigidity coefficients are modified. If this active contribution decreases the effective values $\gamma$ or $\kappa$, in which case, $p_a$ would make the interface less stable for any mode.  

When $R_0^2 > 5/2$, there are only two possibilities: the equation above has no positive root; or, there is a single positive root given by  
\begin{gather}\label{eq:mmax}
 \mm^2\,= \,\frac{3}{4}-\frac{3 R_0^2}{10}+ \left[\left( \frac{3}{4}-\frac{3 R_0^2}{10}\right )^2+\frac{2}{5}(\alpha -\beta)\,R_0^5 + \frac{1}{5}\,R_0^2 -\frac{3}{10} \right]^{1/2}\,.
 \end{gather}
It can be further shown that the root above corresponds to a maximum. We see that $\mm$ depends on the driving parameters only through the factor $\alpha - \beta$, or equivalently, through the velocity,
\begin{gather}
 \delta \,= \,(\alpha - \beta) R_0 \,=\, \frac{\Delta \eta }{\bar \eta}( v_{\mu} + v_q)\, +\, (\frac{\Delta \eta}{\bar \eta }+1)\, \frac{ v_a}{2 }\,.
 \label{eq:delta}
 \end{gather}
 
Although $\mm$ depends only on $\delta$, conditions for instability depend on both $\alpha$ and $\beta$. When the inner domain is large therefore, if $\mm \leq 1$ or if no real $\mm$ exists, $\omega$ decreases monotonically in the domain of interest ($m \geq 1$). The other possibility is that a maximum in $\omega$ occurs at an $\mm > 1$. For each of these cases, there are two alternatives. For monotonically decreasing $\omega$, the mode $m = 1$ is least stable. Stability then is determined by $\omega$ at $m = 1$. As noted earlier, $\omega_1 = (m = 1) = -\beta$; hence, systems with monotonically decreasing $\omega$ are stable when $\beta \geq 0$, and unstable otherwise. When $\omega$ has a maximum at an $\mm > 1$, a system is stable if $\omega_\mathrm{max} = \omega(\mm) \leq 0$, and unstable otherwise. There are thus four qualitative possibilities:
\begin{enumerate}[label=(\roman*)]
\item{ a stable interface with monotonically decreasing $\omega$ for all  $m \geq 1$, with $ \omega_1 \leq 0$;}
\item{ an unstable interface monotonically decreasing $\omega$ for all  $m \geq 1$, with $\omega_1  > 0$;}
\item{ a stable interface with a maximum in the dispersion curve such that $\omega_\mathrm{max}\leq  0$;}
\item{ an unstable interface with a maximum in the dispersion curve such that $\omega_\mathrm{max} > 0$.}
\end{enumerate}

\begin{figure}
     \begin{subfigure}{0.49\textwidth}
               \centerline{\includegraphics[width = 0.90\textwidth]{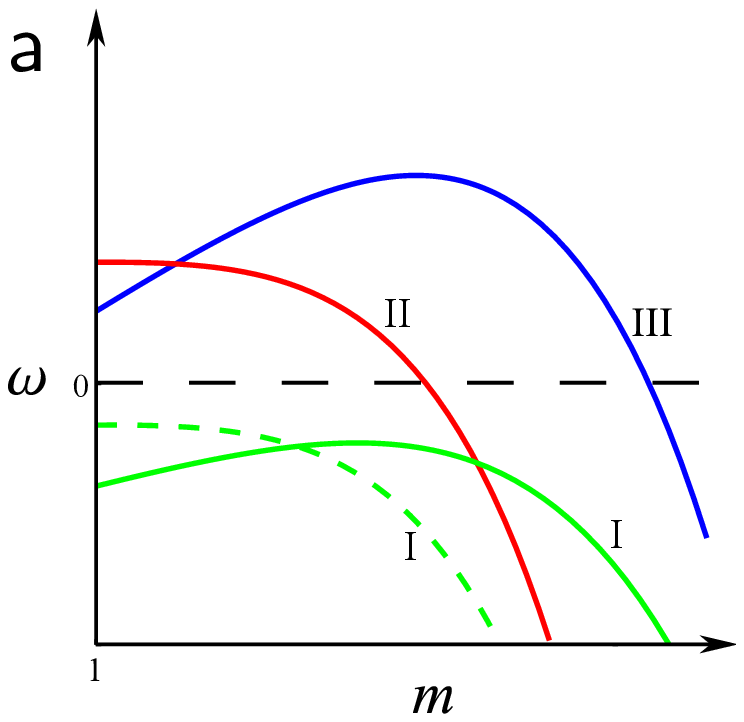}}
     \end{subfigure}\hfill
    \begin{subfigure}{0.49\textwidth}
             \centerline{ \includegraphics[width =0.90 \textwidth]{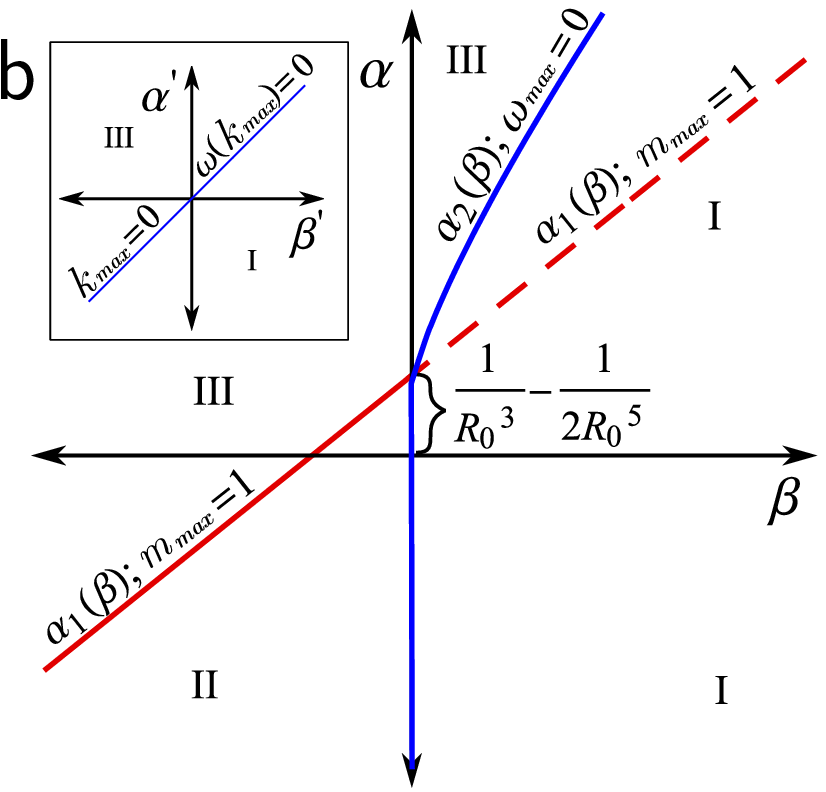}}
   \end{subfigure}
\caption{\label{fig:combined} (a) Typical dispersion curves for ({I}) stable interfaces (green continuous and dashed curves), (II) unstable interfaces with $m = 1$ as the most unstable mode (red curve), and (III) unstable interfaces with $m >1$ for the most unstable mode (blue curve).  (b) Sketch of state diagram in parameter space; $\alpha_1(\beta)$ and $\alpha_2(\beta)$ are given by eqns. \eqref{eq:alpha_1} and \eqref{eq:omega(m)} respectively. The inset in (b) shows the stability regimes for flat interface, with $\alpha' = \alpha R_0$, and $\beta' = \beta R_0$. }
\end{figure}


Figure~\ref{fig:combined} (a) shows these four typical shapes of dispersion curves. From a stability perspective,  the two cases where systems are stable (green continuous and dashed curves in Fig. ~\ref{fig:combined} (a)) are qualitatively indistinguishable, irrespective of whether the dispersion curve has a maximum or not. On the other hand, the behaviour of an unstable system  with the red curve in Fig.~\ref{fig:combined} (a) is qualitatively different from one that is is described by the blue curve. The most unstable mode in the former case is the translational mode, $m = 1$. If this mode grows and dominates, the interface can be expected to deform into an asymmetric shape. \citet{blanch2013spontaneous} have shown that such shape anisotropy can also generate net translation. On the other hand, if the translational mode is most unstable, the interface may not develop distinct fingers: the whole domain is effectively a single finger. In contrast, when the dispersion curve has a distinct maximum at $\mm > 1$, a clear fingering instability develops.

Three distinct stability states can thus be identified:
\begin{enumerate}[label=\Roman*]
\item -- stable interface, with a dispersion curve that either decreases monotonically,  or has a stable maximum;
\item -- unstable interface, with a dispersion curve that decreases monotonically, with the most unstable mode at $m = 1$;
\item -- unstable interface, with a dispersion curve that has a maximum at $m > 1$.
\end{enumerate}
These states are determined by $R_0$, $\alpha$ and $\beta$. For any given $R_0$, the states can be represented as regions in parameter space defined by $\alpha$ and $\beta$. Figure~\ref{fig:combined} (b) shows these regions and their boundaries, which are determined as follows.

From eqn.~\eqref{eq:mmax}, it is clear that $\mm$ increases with $\alpha$. Therefore, a physically observable maximum in the dispersion curve when $\alpha$ is larger than the value $\alpha_1$ at which $\mm = 1$. Substituting $\mm = 1$ in eqn.~\eqref{eq:mmax} therefore, we obtain
\begin{gather}\label{eq:alpha_1}
\alpha_1= \beta  + \frac{1}{R_0^3} -\frac{1}{2R_0^5}\,.
\end{gather}  
Any system  point above this line in the $\alpha-\beta$ space, that is for any $\alpha > \alpha_1$, a system is stable at any $\beta$  if $\omega_\mathrm{max} \leq 0$.  Substituting $m = \mm$ in eqn.~\eqref{eq:omega(m)} from eqn.~\eqref{eq:mmax},  we can solve  $\omega_\mathrm{max} = 0$ for $\alpha$ at any given $\beta$.  Using a continuation approach, the solution $\alpha_2(\beta)$ can be shown to be given by the differential equation,
\begin{gather}\label{eq:dalpha2bydbeta}
\d{\alpha_2}{\beta}=-\pd{\omega_\mathrm{max}/\partial \beta}{\omega_\mathrm{max}/\partial \alpha}=\frac{\mm}{\mm-1}\,.
\end{gather}
This equation can be integrated with respect to $\beta$. The initial condition is obtained by noting the $\alpha_1(\beta)$ and $\alpha_2(\beta)$ must intersect at $\alpha_2=\alpha_1=1/R_0^3 -1/2R_0^5$ and $\beta=0$, at (from eqn. \eqref{eq:alpha_1}). It is found that real values for $\alpha_2$ exist only for $\beta \geq 0$. 

Hence, when $\beta > 0$, the region in Fig. \ref{fig:combined} (b) above the (blue) $\alpha_2(\beta)$ curve represents unstable states that $\omega_\mathrm{max} > 0$ with physically observable $\mm > 1$. Those points correspond to state III. For positive injection rates, points on and below $\alpha_2(\beta)$ have  $\omega < 0$ for all observable modes  and correspond to the stable  state I. As noted earlier,  in systems with a sink at the origin (\textit{i.e.} $\beta < 0$), the translational mode $m = 1$ is already linearly unstable (eqn.~\eqref{eq:omega_m_1}). Therefore, such systems cannot be stable. Further,  there is no real $\alpha_2$ at which $\omega_\mathrm{max} =  0$ with $\mm > 1$.  Consequently, systems with suction either show unstable fingering corresponding to state III  for all $\alpha$ above the  $\alpha_1(\beta)$ (continuous red) curve in Fig.~\ref{fig:combined} (b) or correspond to state II, wherein the interface shape spontaneously becomes asymmetric as $m = 1$ perturbations grow, and the domain may translate as it shrinks in size. 

As $R_0 \rightarrow \infty$, the dispersion relation (eqn.~\eqref{eq:omega(m)}) becomes 
\begin{gather}\label{eq:omegapflat}
\omega(k) \,=\,  -\frac{k}{2} \,\left( \, k^4 \,+\, k^2 \,-\, 2 \,\delta \,\right) \,.
\end{gather} 
Thus, in the limit of a perfectly flat initial interface, stability is governed solely by $\delta$. We see that the interface is marginally stable at $k = 0$, and that perturbations of all wavelengths greater than
\begin{gather}\label{eq:lambda_c}
\lambda_c = \frac{2\sqrt{2} \,\pi}{\left( \, \sqrt{1 + 8\,\delta} - 1 \right)^{1/2}} \,,
\end{gather}
are stable. Hence, flat interfaces are completely stable only if $\delta \leq - 1/8$; if not, a fingering instability is observed. As pointed out earlier, motility at the interface  tends to lead to fingering, irrespective of the relative viscosities of the fluids on either side of the interface. The instability is enhanced by either injection or growth if the outer fluid is more viscous. If the viscosity of the inner fluid is larger on the other hand, a sufficiently large growth rate or injection rate can completely suppress fingering even if the agents at the interface are motile. 

\begin{figure}
     \begin{subfigure}{0.49\textwidth}
                \centerline{\includegraphics[width = 0.90\textwidth]{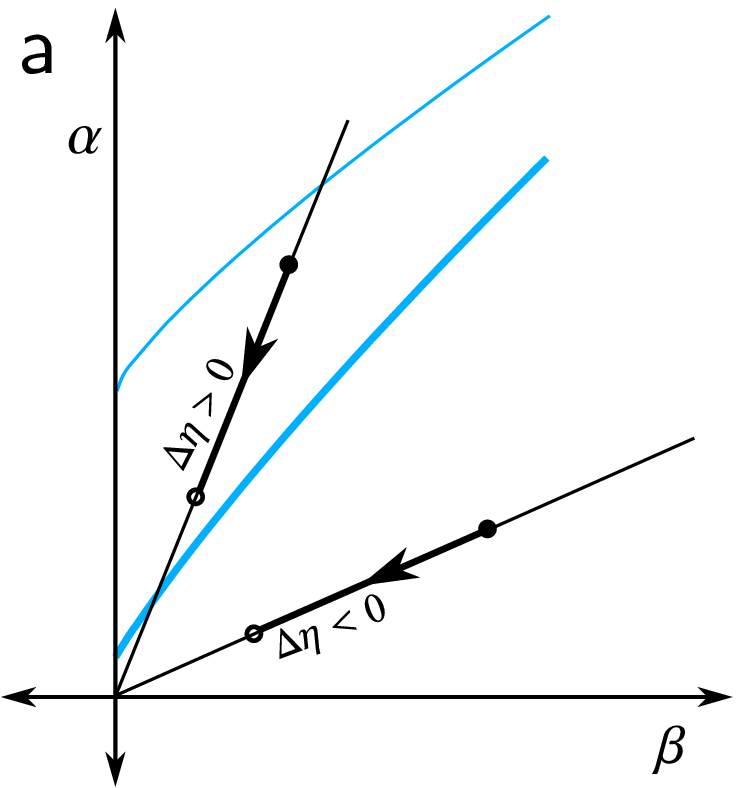}}
     \end{subfigure}\hfill
    \begin{subfigure}{0.49\textwidth}
              \centerline{\includegraphics[width =0.90 \textwidth]{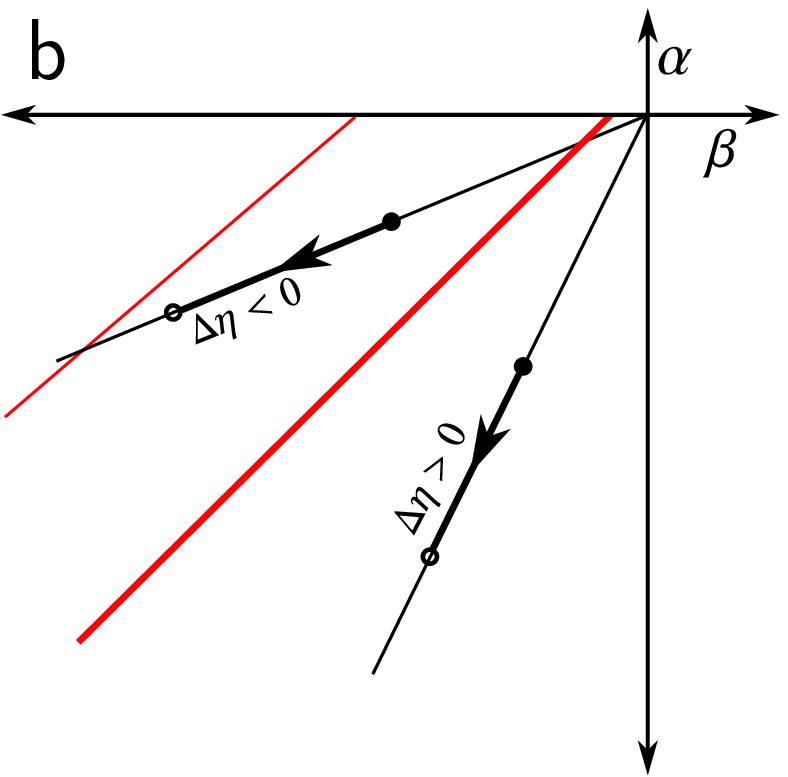}}
   \end{subfigure}
\caption{\label{fig:combined2} Emergence of fingering instability in initially stable interfaces driven purely by a point  source (a) or sink at the origin (b). As the domain size $R_0$ changes, system points evolve from their initial (filled circles) to final (open circles) states along linear operating lines. The stability boundary also changes with $R_0$: initial boundaries - thin coloured curves; final boundaries - thick coloured curves.}

\end{figure}


The results above and the stability diagram can be used to understand the evolution of an initially stable interface with surface tension and bending rigidity. As the inner domain changes in size, the governing parameters $\alpha$ and $\beta$ change with $R_0$, as do the stability boundaries. For example, Fig.~\ref{fig:combined2} demonstrates the emergence of fingering in interfaces driven  purely by a point source or sink.   In this case, the state of any given system evolves on a straight operating line  in parameter space  through the origin with a slope of $(1 - \Delta \eta/\bar \eta) = \eta_1/\bar \eta$. On any such operating line, system points approach the origin  as $R_0$ increases  for a point source ($\beta > 0$; Fig.~\ref{fig:combined2} (a)), and move away from the origin as $R_0$ decreases for a point sink ( $\beta < 0$;  Fig.~\ref{fig:combined2} (b)).   An unstable interface is possible if the operating line intersects the stability boundary.  For a point source, the stability boundary (blue curve in Fig. .~\ref{fig:combined2} (a)), $\alpha_2 (\beta)$, approaches the line of unit slope passing through the origin, as $R_0 \rightarrow \infty$. Since its intercept  on the $\alpha$-axis scales as $R_0^{-3}$ when $R_0 \gg 1$, we see that it can overtake an initially stable system point  moving on an operating line of slope larger than unity \textit{i.e.} when $\eta_1 < \bar \eta$. For a point sink, on the other hand, systems can become critical when $\eta_1 > \bar \eta$.

Any system with $q = 0$, lies on the vertical  axis of the state space diagram. An example of such an interface is a growing domain of cells or a tissue layer. To the extent that such domains can be considered as being viscous fluids, in the absence of any motility at the interface, $\alpha = (\Delta \eta/ \bar \eta)\,\mu/2$ in such systems. If the positive growth rate $\mu$ is constant, the system state point on the $\alpha$ axis does not change. As in the case of a point source discussed above, the intercept of the stability curve on the $\alpha$-axis shifts towards the origin, and fingers emerge once the domain grows beyond a critical size, $R_{0,\,\mathrm{c}} \sim \mu^{-1/3}$. The interface is stable if the growing inner fluid has the larger viscosity. 

Another interesting example where $\beta = 0$ is the case of lamellar cell fragments \citep{verkhovsky1999self}. These are fragments of single eucaryotic cell created such that the cytoskeleton lacks myosin motors. The fragments are observe to migrate across surfaces driven solely by actin polymerization at the interface. Actin filaments are depolymerized continually in the interior.  \citet{callan2008viscous} and \citet{blanch2013spontaneous} suggested that these fragments can be modeled as viscous fluids with actin depolymerization accounted for by a (negative) growth rate that balances the outward active velocity at the interface due to actin polymerization \textit{i.e.} $\mu = - v_a/ (2 \,R_0)$. In this case, $\alpha = v_a/ (2 R_0)$. For large fragments, the interface undergoes a fingering instability if its mean size is larger than $R_{0,\,\mathrm{c}}  \sim v_a^{-1/2}$, irrespective of the viscosity contrast. \ Nonlinear stability analysis shows that, in the absence of bending resistance, symmetry breaking and spontaneous motility emerge when fingers grow to finite size \citep{blanch2013spontaneous}.
 
In conventional systems driven purely by injection at the origin, it has been shown in simulations and experiments that the number of fingers can be kept fixed as the inner domain expands by using a time-dependent $q$. This can in principle be extended more generally to active interfaces. Equation~\eqref{eq:mmax} can be reorganized to show that for large domains, $\delta \sim R_0^{-2}$, in order for $\mm$ to remain constant. In the case of pure injection, this leads to $q \sim R_0^{-1}$. This can be shown to be satisfied if the injection rate varies as $q \sim t^{-1/3}$. Similar relationships can  be derived for other driving modes.

\section{\label{s:conclusions} Conclusions}

We have shown that the original analysis of  \citet{saffman1958penetration} provides a unified approach for understanding  interfacial morphologies in a wide range of active systems. The dependence of stability conditions  and initial size of  fingers on fluid, interfacial and driving parameters can be described in terms of just two dimensionless parameters. The instability predicted in lamellar cell fragments driven by actin polymerization and treadmilling \citep{callan2008viscous, blanch2013spontaneous} is observed to be a special case of more general active interfaces. The analysis here could provide the basis for understanding how confined biological systems may act to regulate the morphologies that emerge at their interfaces. An interesting phenomenon where fingering instability may be important is the formation of distinctive finger-like rafts at the edge of confined monolayers of motile bacteria that advance by furrowing through a thick agar substrate \citep{gloag2013self, zacherson2017network}. Our analysis is also relevant for cytoskeleton-driven motility in single eucaryotic cells \citep{kabaso2011theoretical} or tissue monolayers  \citep{mark2010physical}. The results above further show that it might be possible to combine injection and activity to manipulate interfacial morphology.


\end{document}